\documentclass[smallextended]{svjour3}  % Comment this line 
\usepackage[numbers,sort&compress]{natbib}

\usepackage{hyperref}
\usepackage{graphicx}  % for pdf, bitmapped graphics files
\usepackage{qcircuit}
\usepackage[dvipsnames]{xcolor}\fboxsep22pt
\usepackage{amsmath,amsfonts,bm}
\usepackage[left=2cm,right=2cm,top=2cm,bottom=2cm]{geometry}
\usepackage{colortbl}

\newcommand{\ket}[1]{\ensuremath{\left|#1\right\rangle}} % Dirac Kets
\renewcommand{\bf}[1]{\ensuremath{\mathbf{#1}}}

\begin{document}

\title{Combinatorial optimization through variational quantum power method
}

%\titlerunning{Short form of title}        % if too long for running head

\author{Ammar Daskin
}

%\authorrunning{Short form of author list} % if too long for running head

\institute{A. Daskin \at
              Dept. of Computer Engineering \\ 
              Istanbul Medeniyet University, Istanbul, Turkey
              \email{adaskin25@gmail.com}           %  \\
}

\date{Received: date / Accepted: date}
% The correct dates will be entered by the editor

\maketitle

%%%%%%%%%%%%%%%%%%%%%%%%%%%%%%%%%%%%%%%%%%%%%%%%%%%%%%%%%%%%%%%%%%%%%%%%%%%%%%%%
\begin{abstract}
The power method (or iteration) is a well-known classical technique that can be used to find the dominant eigenpair of a matrix. 
Here, we present a variational quantum circuit method for the power iteration, which can be used to find the eigenpairs of unitary matrices and so their associated Hamiltonians.
 We discuss how to apply the circuit to combinatorial optimization problems formulated as a quadratic unconstrained binary optimization and discuss its complexity. 
 In addition, we run numerical simulations for random problem instances with up to 21 parameters and observe that the method can generate solutions to the optimization problems with only a few number of iterations and the growth in the number of iterations is  polynomial in the number of parameters. Therefore,  the circuit can be simulated on the near term quantum computers with ease.
 
\end{abstract}

%%%%%%%%%%%%%%%%%%%%%%%%%%%%%%%%%%%%%%%%%%%%%%%%%%%%%%%%%%%%%%%%%%%%%%%%%%%%%%%%
\section{Introduction}

Combinatorial optimization \cite{korte2012combinatorial} is the process of choosing a combination of parameters which gives the best solution for a problem described in a discrete domain.
In particular, we describe the optimization as a process of finding a binary vector $\bf{x}$ which gives a minimum of the following objective function described in the general quadratic form:
\begin{equation}
f(\bf{x}) = \sum_ic_ix_i + \sum_{i<j} q_{ij}x_ix_j,
\end{equation}
where $\bf{x_i}$ is a binary vector element which represents the value of the parameter $i$ in the optimization. 
$c_i$ and $q_{ij}$ are real-valued coefficients. 
This optimization can be applied to many NP-hard problems such as  the set cover,  max-cut, traveling salesman, and facility scheduling problems and the integer programming.
Because the parameters represented by $\bf{x_i}$s can be easily mapped onto the spin operators, this optimization is also heavily studied on quantum computers by using especially the adiabatic quantum computing \cite{farhi2000quantum, albash2018adiabatic}. In this mapping, we simply put Pauli spin operators in place of the parameters:
\begin{equation}
\label{eqHamiltonian}
\mathcal{H} = \sum_ic_i\sigma_z^{(i)} + \sum_{i<j} q_{ij}\sigma_z^{(i)}\sigma_z^{(j)}
\end{equation}
While an alignment of the spin operators in the above Hamiltonian formulation describes a feasible combination of the parameters, 
the eigenvalue obtained by this alignment describes the fitness value of the objective function.
Therefore, in the minimization problem, the alignment that produces the lowest eigenvalue of the Hamiltonian gives the solution of the combinatorial problem. 
Here, note that while the parameter $x_i\in\{0,1\}$,
 the alignment for $\sigma_z^{(i)}$ can be in $\{-1, 1\}$. 
 
Mapping the spin operators to the qubits and using one- and two-qubit quantum rotation gates, the Hamiltonian in Eq.\eqref{eqHamiltonian} can be used on the standard quantum computers. 
This leads to a unitary matrix described as $\mathcal{U} = e^{i\mathcal{H}}$.
If $\mathcal{H}$ has the eigenvalues $\lambda_j$, where $j = 0, 1, \dots$, then the eigenvalues of $\mathcal{U}$ comes in the form  $e^{i\lambda_j}$.
As a result, the optimization becomes finding  the eigenvalue $e^{i\lambda_j}$ of $\mathcal{U}$  with the minimum phase value: i.e. the lowest eigenvalue $\lambda_j$ of $\mathcal{H}$.

\begin{figure*}[t]
	\begin{center}
\colorbox{SpringGreen!50}{
\mbox{
			\Qcircuit @C=0.5em @R=.5em {
				&	 \gate{R_z(c_0)} 	&	\qw	&	\ctrl{1}	&	\ctrl{2}	& \ctrl{3}	&	\qw	&	\qw	&	\qw	&\qw	\\	
				&	 \gate{R_z(c_1)} 	&	\qw	&	\gate{R_z(q_{01})}	&	\qw	&	\qw	&	\ctrl{1}	&	\ctrl{2}	&		\qw&\qw	\\
				&	 \gate{R_z(c_2)} 	&	\qw	&	\qw	&	\gate{R_z(q_{02})}	&	\qw	&	\gate{R_z(q_{12})}	&	\qw	&		\ctrl{1}&\qw	\\
				&	 \gate{R_z(c_3)} 	&	\qw	&	\qw	&	\qw	&	\gate{R_z(q_{03})}	&	\qw	&	\gate{R_z(q_{13})}	&	\gate{R_z(q_{23})}&\qw	\\
			}
		}}
	\end{center}
	\caption{\label{FigU}The circuit for the Hamiltonian in Eq.\eqref{eqHamiltonian} that generates the equivalent $\mathcal{U}$ which is a diagonal matrix with the elements $e^{i\lambda_j}$ where $\lambda_j$s represent the objective function evaluations for the solution space of the optimization problem. 
The eigenvectors represent the solution space by indicating the different combinations of the parameters.
	}
\end{figure*}
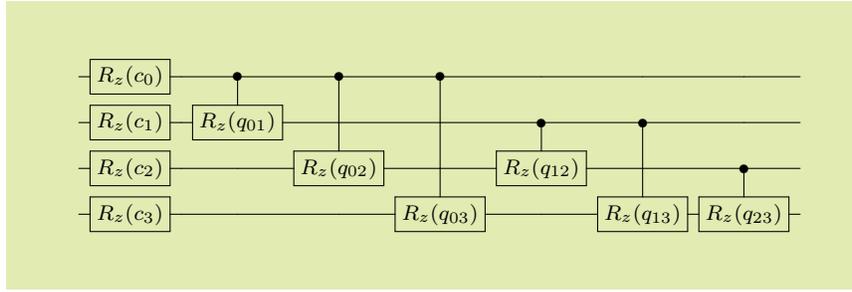

To map a given $\mathcal{H}$ or $f(\bf{x})$ to a quantum circuit,  as done in Ref.\cite{daskin2018qpm}, for each single term $c_ix_i$ we use the following rotation gate on the $i$th qubit:
\begin{equation}
R_z(c_i) = \left(\begin{matrix}
1 & 0\\
0&e^{ic_i}
\end{matrix}\right).
\end{equation}
For the terms $q_{ij}$,  the above quantum gate with the angle value $q_{ij}$ is put on the $j$th qubit and controlled by the $i$ qubit. 
The resulting circuit, $\mathcal{U}$, is illustrated in Fig.\ref{FigU} for 4 qubits: For an $n$-parameter optimization, the circuit requires $n$ qubits (the circuit width) and $n^2$ number of quantum operations (the circuit depth). 

In quantum computing, the phase estimation algorithm is the standard process  to find the eigenvalue of a unitary matrix. This algorithm requires an approximated initial eigenvector which determines the convergence of the algorithm: i.e. the probabilities of the states in the output are related to the overlaps of the eigenstates to the initial state. Therefore, in this algorithm a good approximate initial vector close to the desired eigenstate should be provided. This makes difficult to apply this algorithm when such initial state is not available.
   
The quantum version of the power method \cite{daskin2018qpm} similar to the classical power does not require an initial eigenvector estimate and can be used instead of the phase estimation algorithm. However, as explained in the next section, the required number of qubits in the implementation  of this method on quantum computers grows with the number of iterations which may be exponential in the case where the eigengap is too small. 
This particularly hinders the implementation of this method  on the near term quantum computers where the number of qubits limited.
In Sec.\ref{SecVQPM}, we introduce a variational quantum circuit for the power method, which can be implemented with ease on near term quantum computers. 
In Sec.\ref{SecCombOpt}, we show how to employ this circuit for the combinatorial optimization and discuss its complexity. Then, we present the result of the numerical simulations for random problem instances  with up to 21 parameters and discuss some possible future directions.

\section{Quantum power method}
The quantum version of the classical shifted-power iteration is described in Ref.\cite{daskin2018qpm}:
For a given initial vector $\bf{v_0}$, the classical iterative algorithm can be simply described  by a matrix-vector transformation:
 \begin{equation}
 \bf{v_k} = \frac{\mathcal{H}v_{k-1}}{||\mathcal{H}v_{k-1}||}.
 \end{equation}
The above standard version of the power method can be implemented on quantum computers by using $\mathcal{H}$ in a circuit extended with some auxiliary qubits. In that case the complexity would increase  by the required ancilla qubits. 
One may consider using $\mathcal{U}$ instead of $\mathcal{H}$; however, because  $\mathcal{U}$ is unitary, the algorithm would not converge.
  
To solve this convergence problem,  we consider the matrix $(I+\mathcal{U})$ where $I$ is the identity matrix.  
While the eigenvectors of this matrix are the same as the ones of $\mathcal{U}$, the eigenvalues come in the form $(1+e^{i\lambda_j})$. 

If the eigenvalues of $\mathcal{H}$ are  real and ordered as $|\lambda_0|\leq |\lambda_1| \leq \dots$ where $\lambda_j \in [-\pi/2,\pi/2]$, then the magnitudes of the eigenvalues of $\mathcal{U}$, i.e. $|1+e^{i\lambda_j}|$s are ordered as:
\begin{equation}
(2+2\cos\lambda_0)  \geq (2+2\cos\lambda_1) \geq \dots
\end{equation}
For simplicity, we will assume $\lambda_j \in [0,\pi/2]$ in the rest of the paper.

Since the power method would converge to the principal eigenvalue (the one with the largest magnitude), the above order shows that the algorithm for $(I+\mathcal{U})$ would converge the eigenvector associated with the eigenvalue $(2+2\cos\lambda_0)$.  
Therefore, using $(I+\mathcal{U})$ we can estimate the eigenvector of $\mathcal{H}$ associated with its smallest eigenvalue $\lambda_0$.

\subsection{Quantum circuit implementation}

\begin{figure}[t]
	\begin{center}
\colorbox{SpringGreen!50}{
\mbox{
		\Qcircuit @C=0.7em @R=.7em {
\lstick{\ket{0}} & \gate{H} &  \qw &\ctrl{1} & \gate{H} & \meter& \cw &\cw &  P_0\\
\lstick{\ket{\bf{v_{k}}}} &\qw & \qw  &\gate{\mathcal{U}}  & \qw & \rightarrow {\frac{(I+\mathcal{U})\ket{\bf{v_{k}}}}{||(I+\mathcal{U})\ket{\bf{v_{k}}}||}}
			}}}
	\end{center}
	\caption{\label{FigQPM} An iteration of the quantum power method.}
\end{figure}
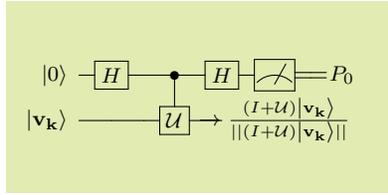

Assume that we are given the initial state \ket{0}\ket{\bf{v_k}} as depicted in Fig.\ref{FigQPM}.
If we apply the Hadamard gate to the first qubit, we obtain
\begin{equation}
\frac{1}{\sqrt{2}}\left(\ket{0}\ket{\bf{v_k}} + \ket{1}\ket{\bf{v_k}}\right).
\end{equation}
The application of the controlled $\mathcal{U}$ leads to the following quantum state:
\begin{equation}
\frac{1}{\sqrt{2}}\left(\ket{0}\ket{\bf{v_k}} + \ket{1}\mathcal{U}\ket{\bf{v_k}}\right).
\end{equation}
 The second Hadamard gate on the first qubit transforms the state into:
 \begin{equation}
\frac{1}{2}\left(\ket{0}\ket{\bf{v_k}} + \ket{1}\ket{\bf{v_k}}+
\ket{0}\mathcal{U}\ket{\bf{v_k}} - \ket{1}\mathcal{U}\ket{\bf{v_k}}
\right),
\end{equation}
which can be written more concisely as:
\begin{equation}
\frac{1}{2}\left(\ket{0}\left(I+\mathcal{U}\right)\ket{\bf{v_k}} + 
\ket{1}\left(I-\mathcal{U}\right)\ket{\bf{v_k}}\right).
\end{equation}
 Measuring the first qubit in \ket{0} state, the final quantum state collapses onto the following state:
\begin{equation}
\frac{\left(I+\mathcal{U}\right)\ket{\bf{v_k}}}{||\left(I+\mathcal{U}\right)\ket{\bf{v_k}}||}.
\end{equation} 
 The above is the same as the result of an iteration in the power method. Repeating this circuit with the collapsed output state and the same operations with another control qubit  is equivalent to the classical power method. Hence, the final collapsed output state would converge on to the eigenvector of  $\mathcal{U}$.
  
Here, in each iteration, the probability of measuring $\ket{0}$ on the first qubit is defined as
  \begin{equation}
  P_0 =\frac{ \left\Vert \left(I+\mathcal{U}\right)\ket{\bf{v_k}}\right\Vert^2}{4}.
\end{equation}   
Since the eigenvalues of $(I+\mathcal{U})$ are in the form $(2+2\cos\lambda_j)$ and for $\lambda_j \in [0, \pi/2]$, $\cos\lambda_j$ can be maximum 1 and minimum 0; we observe
\begin{equation}
2\leq \left\Vert \left(I+\mathcal{U}\right)\ket{\bf{v_k}}\right\Vert^2  \leq 4.
\end{equation}
Therefore,
\begin{equation}
0.5 \leq P_0\leq 1.
\end{equation} 
Note that for $\lambda_j \in [0,1]$ it is guaranteed that $P_0\geq 0.77$. Therefore, it is always easy to collapse the quantum state onto the state where the first qubit is in \ket{0}. 
 Also note that when $P_0$ is maximized, the power iteration converges to the minimum eigenvalue of $\mathcal{U}$.
 
 The required number of iterations in the power method is related to the eigengap. For a matrix with an eigengap $\gamma$, the principal eigenvector and eigenvalue can be found by using $O(1/\gamma)$ number iterations.
 In the quantum case, since in each iteration we use a new qubit for the controlled operations, the algorithm would require $O(1/\gamma)$ number of operations and qubits.
 This would hinder the implementation of the method on the near term quantum computers which have limited coherence times and can employ a small number of qubits.
In the next section, we will describe a variatonal version of this method which can be used easily with the near term quantum architectures.
 
\section{Variational quantum power method (VQPM)}
\label{SecVQPM}
\begin{figure}[t]
	\begin{center}
\colorbox{SpringGreen!50}{
\makebox[0.4\textwidth]{
		\Qcircuit @C=0.7em @R=.7em {
\lstick{\ket{0}} & \gate{H} &  \qw &\ctrl{1} & \gate{H} & \meter& \cw &\cw &  P_0\\
\lstick{\ket{0}} &\qw & \gate{R_y({\theta_0})}  &\multigate{2}{\mathcal{U}}  & \qw & \qw & \meter&\cw &\theta_0\\
\lstick{\ket{0}} & \qw & \gate{R_y({\theta_1})}  &\ghost{\mathcal{U}}  &\qw & \qw &\meter &\cw& \theta_1\\
\lstick{\ket{0}} & \qw & \gate{R_y({\theta_2})}  &\ghost{\mathcal{U}}  & \qw & \qw &\meter &\cw& \theta_2\\
			}}}
\\
\colorbox{Turquoise!20}{
\begin{minipage}{.4\textwidth}
\textbf{Optimization:} 
find $\bf{\theta}$ that maximizes $P_0$ and so gives the minimum $\lambda_j$.	
\end{minipage}}

	\end{center}
	\caption{\label{FigVQPM}Quantum circuit used to evaluate the objective function.$P_0$ is the probability of measuring $\ket{0}$ on the first qubit. $\theta_i$ is the angle value for the rotation gate that gives the probabilities observed on the $i$th qubit in the collapsed state. 
The circuit is iterated until $P_0$ is maximized or we measure \ket{0} or \ket{1} with some certainty on the remaining qubits.}
\end{figure}
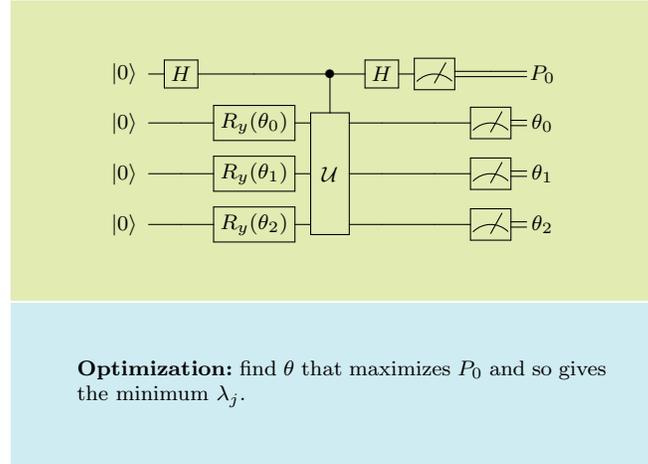

In the variational version of the quantum power method which is depicted in Fig.\ref{FigVQPM},  we start with $\ket{0}\ket{\bf{0}}$. 
Before the application of the controlled $\mathcal{U}$, we apply $R_y(\theta_0)\otimes R_y(\theta_1) \otimes \dots$: Here,
\begin{equation}
R_y(\theta_j) = \left(\begin{matrix}
\cos(\theta_j) & -\sin(\theta_j)\\
\sin(\theta_j)& \cos(\theta_j).
\end{matrix}\right)
\end{equation}
The value of $\theta_j$s are determined from the individual qubit measurement done in the previous iteration of the algorithm.
As in the standard variational quantum algorithm \cite{peruzzo2014variational},  a classical optimization algorithm such as the Nelder–Mead method \cite{nelder1965simplex} can be employed   to maximize the probability $P_0$ by adjusting the values of $\theta_j$s in the subsequent iterations of the optimization.

\section{Using VQPM in combinatorial optimization}
\label{SecCombOpt}
Let us assume we are given $\mathcal{U}$ whose  eigenpairs represent the solution domain of a combinatorial problem.
In the circuit, each qubit in the second register (the first register consists of the control qubit) is associated with a parameter of the optimization problem.
The change in the value of any parameter, i.e. being 0 or 1, changes the fitness value of the objective function.
In the power iteration, we can easily observe that each qubit in the second register gradually converges to either $\ket{0}$ or $\ket{1}$ state: i.e. the either probability converges to 1.

Since the power iteration converges toward the principal eigenvalues, this convergence is faster for those parameters whose values differentiate the eigenvalues that are smaller:
For the parameters whose values differentiate the eigenvalues $\lambda_j >> \lambda_k$ converges more easily than those whose values differentiate the eigenvalues that are very close to each other.

 \subsection{Precision in the measurement}
In classical systems, one single measurement on the bit is sufficient to  obtain the single bit of information.
However, a qubit in quantum systems can be represented by a point on the Bloch sphere. Therefore, the p-copies of a quantum system provide $\log_2(p+1)$ bits of information about the state of the qubit. 
In other words, given p-copies of a quantum system, we can determine the state of a qubit with an accuracy greater than or equal to $\frac{2}{\sqrt{p+1}}$ \cite{christandl2012reliable}.
In quantum computers, the accuracy in the measurement of a quantum state of a qubit is fundamentally limited by the statistical principles of the quantum mechanics \cite{cramer1999mathematical,braunstein1994statistical} and changes based on the properties (e.g. the fidelity of the implemented quantum gates) of the underlying quantum computer technology \cite{hou2016achieving}. 
This accuracy can be improved by using an appropriate data analysis procedure along with the quantum state tomography (e.g. see \cite{christandl2012reliable, hentschel2010machine} ).

Here note that the accuracy in the measurement can guarantee a bound on the error of the solution we obtain at the end of the variational circuit:
 If we have a precision $\gamma$ in the measurement, then we can expect the fitness value of the obtained solution to differ from the best value by at most  $O({\gamma})$.

\subsection{Convergence}
In the classical power iteration, the convergence is determined by the eigegap. 
In VQPM, every iteration we measure each qubit and check to see if the probability difference on any qubit has reach a certain point indicated by a parameter, $p_{diff}$. 
If this parameter is small, then we expect the algorithm to converge onto a solution very fast. 
However, if it is too small (e.g., very close to the precision of the quantum computer or the measurement), it leads the algorithm to converge one of the non-optimal solutions.
Therefore, before running the algorithm for a real world problem, this parameter should be optimized with some test problems and determined by considering the underlying quantum computer.  
Fig.\ref{FigExample} shows how the change in this parameter  affects the required number of iterations for the example problem described in the next section.   

\subsection{Complexity}
The total complexity is determined by the number of iterations. Each application of the circuit requires $O(n^2)$ simple quantum gates on $n+1$ qubits. To determine the probabilities for each qubit, if we use $p$ number of copies of the circuit, then in each iteration we apply $O(pn^2)$ quantum operations. In general, the precision and the number of iterations are closely correlated: if we can increase the precision, then we can decrease the required number of iterations.   

\section{Numerical Example and Random Experiments\protect\footnote{ You can access the simulation code for the variational quantum power method for random quadratic binary optimization from the link: 
\href{https://github.com/adaskin/vqpm\_public}{https://github.com/adaskin/vqpm\_public}}}

For $x_i \in \{0, 1\}$, where $i = 0, \dots, 3$; 
consider the following random quadratic unconstrained binary minimization problem:  
\begin{equation}
\begin{split}
y =\ & 4.02377326x_0  +1.4338403x_1 -3.60973431x_2 -0.52469588 x_3\\ 
 & -1.06286586 x_0 x_1  +  0.49009314 x_0 x_2 +0.95332512 x_0 x_3  \\
&-1.4136876 x_1 x_2  + 0.29605018 x_1 x_3 
  -0.7966874 x_2 x_3.  
\end{split}
\end{equation}
From the above equation, we can derive the following coefficient matrix:
\begin{equation}
Q = \left( \begin{matrix}
4.02377326 &-1.06286586 &0.49009314 &0.95332512 &\\
0 &1.4338403 &-1.4136876 &0.29605018 &\\
0 &0 &-3.60973431 &-0.7966874 &\\
0 &0&0 &-0.52469588 &\\
\end{matrix}\right)	
\end{equation}
Then, the optimization problem becomes: $\min{\bf{x}^TQ\bf{x}}$, where $\bf{x}^T$ is the transpose of the vector $\bf{x}$ (Note that $Q$ is given in an upper-diagonal form to make the quantum circuit simpler.). Many NP-hard problems can be mapped into the above form as a quadratic optimization problem.  We refer the reader to  Ref.\cite{glover2019quantum} which illustrates such mappings of different problems.

In our simulations, we first  mapped the solution space of the optimization problem into the region $[-\pi/4, +\pi/4]$ by making the following scaling: 
\begin{equation}
\begin{split}
Qscaled = & \frac{Q}{\sum_{i,j}|Q_{ij}|}\times \frac{\pi}{4}
\\ = & \left( \begin{matrix}
0.216386 &-0.05715762 &0.02635568 &0.05126686 &\\
0 &0.07710747 &-0.07602372 &0.01592066 &\\
0 &0 &-0.19412028 &-0.04284337 &\\
0 &0 &0 &-0.02821651 &\\
\end{matrix}\right)	
\end{split}
\end{equation}

Using the above coefficient in the circuit of Fig.\ref{FigU}, we can generate the solution space of the optimization problem. 
In our implementation, the variational quantum power method converges to the minimum phase in magnitude, i.e. $|\lambda|$. Therefore, to find the minimum in real value, we apply a global phase $e^{i\pi/4}$. This moves the solution space, the $\lambda_j$s in the eigenvalues of $\mathcal{U}$, to the range $[0, \pi/2]$: i.e. new eigenphase value $\tilde{\lambda}_j = \lambda_j+\pi/4$. 
The solution space of the problem is listed in Table \ref{TableSolSpace} with $Q$, $Qscaled$, and after the global phase $\pi/4$.  In the final solution space, the smallest and the second smallest eigenphases are produced with the solution ``0011" and ``0111", respectively. The difference between these values(the eigengap) is  0.017.

\begin{table}\centering
	\caption{\label{TableSolSpace}Table shows the solution space for the example problem described in the text (The highlighted row indicates the optimum solution.). By scaling coeficients and applying a global phase, the phase values, i.e. the value $\lambda_j$ in $e^{i\lambda_j}$, are mapped into $[0, \pi/2]$ range. }
	\begin{tabular}{|l|l|l|l|}
		\hline 
		\rowcolor{Turquoise!20}\hline$x_0x_1x_2x_3$& $y$& $y$ with  $Qscaled$ &  $\frac{\pi}{4}$ added: $\lambda_j  + \frac{\pi}{4}$ \\
		\hline
		0000 & 0.0 & 0.0 & 0.7854 \\
		0001 & -0.5247 & -0.02822 & 0.75718 \\
		0010 & -3.60973 & -0.19412 & 0.59128 \\
		\rowcolor{SpringGreen!50}0011 & -4.93112 & -0.26518 & 0.52022 \\
		0100 & 1.43384 & 0.07711 & 0.86251 \\
		0101 & 1.20519 & 0.06481 & 0.85021 \\
		0110 & -3.58958 & -0.19304 & 0.59236 \\
		0111 & -4.61491 & -0.24818 & 0.53722 \\
		1000 & 4.02377 & 0.21639 & 1.00178 \\
		1001 & 4.4524 & 0.23944 & 1.02483 \\
		1010 & 0.90413 & 0.04862 & 0.83402 \\
		1011 & 0.53607 & 0.02883 & 0.81423 \\
		1100 & 4.39475 & 0.23634 & 1.02173 \\
		1101 & 5.11943 & 0.27531 & 1.06071 \\
		1110 & -0.13858 & -0.00745 & 0.77795 \\
		1111 & -0.21059 & -0.01132 & 0.77407 \\
		\hline
	\end{tabular}
\end{table}

In VQPM as drawn in Fig.\ref{FigVQPM}, we start with $\theta_j = \pi/4$ to generate an equal superposition state on the second register before the application of the controlled $\mathcal{U}$. 
	In the next iteration, we set $\theta_{j}$ to either $\pi/2$ or 0 if there is a probability difference (In our experiments, we observe $\gamma\times 10$ to be a good choice.) between \ket{1} and \ket{0} for the qubit that represents parameter $j$: i.e., we have found the correct alignment for the parameter $j$ at this iteration and so can feed this alignment into the circuit to reduce the search space of the optimization by half.
If the eigengap distinguished by the parameter $j$ is $<\gamma$ through iterations, than the qubit that determines the value of parameter $j$ remains in equal superposition states.
In this case, if we have $m$ parameters whose values are not determined, than the final quantum state indicates the correct eigenvector (the solution to the optimization problem)  with the probability $\geq \frac{1}{2^m}$. 

As discussed in the previous sections, one of the parameters is expected to affect the performance of VQPM in the simulation is the precision of the measurement and the other one is the probability difference between 0 and 1 that is used to decide whether the value of the parameter $x_i$ should be either 0 or 1 in the solution (e.g.  if the probability to see \ket{a} is more than \ket{b} state by the indicated difference, then we prepare this qubit in \ket{a} state for the rest of the iterations and disregard the rotation gate $R_y(\theta)$  on this qubit.). 

\begin{figure}
	\centering
	\includegraphics[width=4.5in]{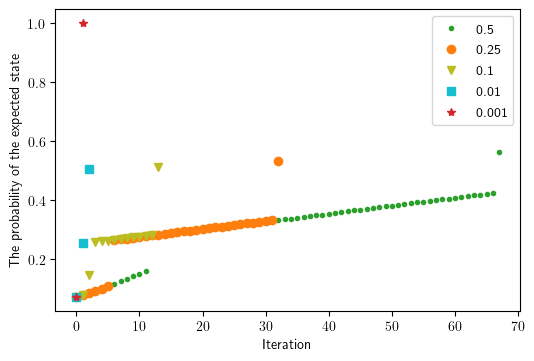}
	\caption{\label{FigExample}The probability change of the expected  state until it reaches  $\geq0.5$ when using the different $p_{diff}$ values to assume a qubit state (i.e. in the measurement if the probability to see the qubit in \ket{a} state is greater than \ket{b} state by $p_{diff}$, then we prepare qubit in \ket{a} state for the rest of the iterations). }
\end{figure}

For the example problem described above; Fig.\ref{FigExample} shows how using  the different $p_{diff}$ values to assume a bit value of a qubit affects the number of iterations required to raise the success probability of the solution state. Although it is accurate to say that we can get the solution state by using a smaller probability difference, it may cause the algorithm to converge on a wrong answer in early iterations. Therefore, one should carefully decide the value of this parameter based on the quantum computer used for the experiment.

In our numerical simulations, we assume that we can measure the state of a qubit with the precision $\gamma = 10^{-4}$ (In the numerical simulation, we round the probabilities to the 4 decimal points.) and assume a qubit should be $1$ or $0$ if the probability difference between qubits greater than 0.001.

\subsection{Generic Random Experiments}

We generate 50 random quadratic binary optimization for each parameter value $n\in [2,21]$.
The random instances are generated by using random $q_{ij}$ and $c_j$ coefficients in Fig.\ref{FigU}. To convert the instances to a minimization problem, we scale the coefficients and apply a global phase $\pi/4$ by using the same way described above for the example problem. Therefore,  VQPM searches for the solution associated with  the minimum phase value. 

\begin{figure}
 \centering
 \includegraphics[width=4.5in]{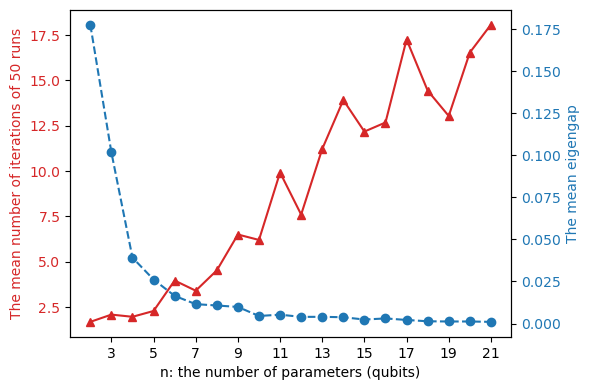}
 \caption{\label{FigRuns}For different $n$ values (the x-axis), the mean number of iterations(the left y-axis) of 50 random quadratic binary optimization instances to reach the success probability $\geq0.5$. The mean eigengap(the right y-axis) in these instances.}
 \end{figure}

\begin{figure}
	\centering
	\includegraphics[width=4.5in]{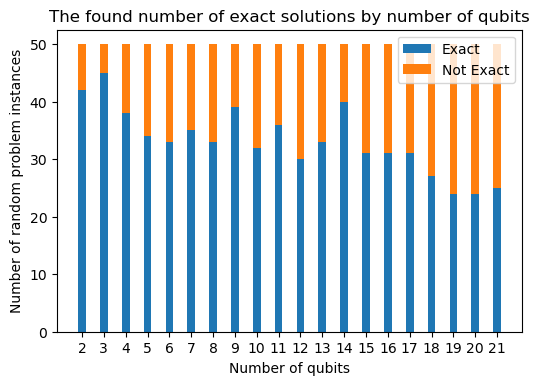}
	\caption{\label{FigExactHist}For different $n$ values (the x-axis), the number of problem instances for which the exact solution is found in 100 iterations.}
\end{figure}

\begin{figure}
	\centering
	\includegraphics[width=4.5in]{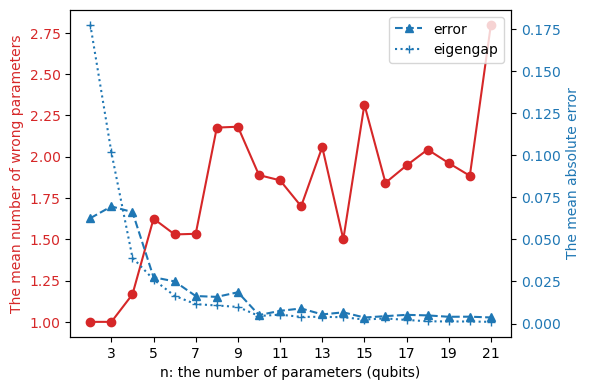}
	\caption{\label{FigMeanError}In the runs where the algorithm converge on non-exact solution;  while on the right, the blue dashed line shows the mean absolute error (i.e. the mean absolute difference to the optimum solution) for different problem sizes, on the left the red line shows the number of parameter values that differ with the optimum solution (i.e. the mean number of 1s in the $XOR$ of the optimum and found solutions).}
\end{figure}

In the numerical simulations, we limit the maximum number of iterations for any problem instances by 100 iterations.
The results of the simulations are presented in Fig.\ref{FigRuns}, Fig.\ref{FigExactHist}, and Fig.\ref{FigMeanError} : 
Fig.\ref{FigRuns} depicts the mean number of iterations for the probability of the eigenvector indicating the solution to reach $\geq 0.5$  versus the number of parameters. It also shows the mean eigengap for each number of parameters. 
As discussed above, when the eigengap is smaller than the precision in the measurement, in our case $\gamma = 10^{-4}$; the alignments of the parameters whose values distinguish the lowest eigenvalues cannot be determined.
In these instances, the output state is composed of such non-distinguishable eigenstates.
This indicates that even if we cannot single out the solution from the huge domain of the optimization problem, we are still able to reduce the problem size by a great deal. In addition, the solution we obtain at the end is guaranteed to be less than the precision, which is $10^{-4}$ in our simulations.
The figure also shows the method is scalable to larger problems since the mean number of iterations grows linearly with the number of parameters.  

Fig.\ref{FigExactHist} shows the number of times the algorithm converges on to an exact and non-exact solutions. As seen from the figure, about half of the time the converged solution is not the exact one. 
This is can be explained by the following reasons:
\begin{itemize}
	\item   As mentioned above,  in some cases, VQPM may converge onto the non-optimum solution by assuming a wrong bit value for a qubit in early iteration. 
	\item At the end of 100 iterations, we have chosen the state with the maximum probability to be the found answer. (Note that in some instances, the optimum solution may still be part of the final state with some probability albeit not the highest.) 
\end{itemize}

In the cases where the algorithm has converged on the non-exact solutions, 
Fig.\ref{FigMeanError} shows how the found solutions differ from the exact solutions: In particular, it depicts the mean numerical difference (the absolute error) and the mean number of different bit values in the found and the exact solutions.  As seen in the figure, the number of parameter values that are estimated wrong is around 2: For about 20 parameters this indicates only a 1 percent error in the estimated parameter values. 
In addition, the absolute error is very close to the eigengap, which is exponentially small in the number of parameters.

\section{Discussion and Future Directions}

In general, the combinatorial optimization problems considered difficult for current classical computers involve more than 50 parameters. Since simulations for this problems on desktop PCs are not feasible for our quantum algorithm (it necessitates storing a quantum state  of $50$ qubits), at this juncture it is not easy to draw conclusion whether the algorithm would outperform classical algorithms. However, the linear growth in the number of iterations and the observed accuracy indicates that the method may be used to in the works to proof ``quantum supremacy".

The real world combinatorial problems generally include some constraint on the solution space. As also shown in Ref.\cite{glover2019quantum}, we can convert a constraint problem to an unconstrained one by adding some penalty terms into the optimization. 
One can also integrate constraint terms as a separate quantum register where a value of the qubit indicates if the constraint is satisfied.
The 0-1 knapsack problem can be considered as an example for a constrained optimization problem \cite{ezugwu2019comparative}: Given a set of $n$ items each of with the value $v_i$ and weight $w_i$, the 0-1 knapsack problem is defined as:
\begin{equation}
\text{maximize }\sum_{i} v_i x_i, \text{ subject to } \sum_{i} w_i x_i \leq W\text{ and } x_i \in \{0,1\}.
\end{equation} 
Note that there is also a more general version which is the quadratic knapsack problem. 
To adjust VQPM in Fig.\ref{FigVQPM}, for values and the weights one can employ two quantum registers: namely the solution  and constraint, represented by $\ket{reg_v}$ $\ket{reg_w}$ respectively. 
 \ket{reg_w} controls whether the candidate solution in \ket{reg_v} satisfies the constraint in the problem. And the result is indicated in qubit \ket{C_0}. 
From this point on, we may follow different paths to integrate the value of  \ket{C_0} into the optimization. 
From this point on, one may control quantum gates on the second register by the first register or apply CNOT gates at the beginning to entangle the qubits in two registers. However, more research is needed to see if this integration could be enough for the optimization.   

Note that  employing $d-$level qubits, one can also use the fractional knapsack problem where we can take some fractions of items. Or more generally one can use this algorithm in integer programming where the value of the items can be any of $d$ distinct integers.

\section{Conclusion}
In this paper, we have presented a variational quantum circuit based on the power method used for the eigendecomposition. We have shown that one can use this circuit for combinatorial optimization problem formulated as an eigenvalue problem. The accuracy of the obtained solution in the circuit is determined by the precision in the measurement of single qubit states. 
In other words, the circuit guarantees that the error in the fitness value of the generated solution for the optimization is less than the precision in the measurement. 
The circuit is simple enough to be used in the near term quantum computers and may be used in the works to prove ``quantum supremacy".  
%%\bibliographystyle{spbasic}
%%\bibliography{ref}

\end{document}